\documentclass[preprint,nofootinbib,amsmath,amssymb,amsfonts,aps,prd,eqsecnum]{revtex4-1}

\usepackage{hyperref}
\usepackage{amsmath}
\usepackage{mathrsfs}

\DeclareMathOperator*{\wlim}{w-lim}

\newtheorem{theorem}{Theorem}

\begin{document}

\title{Examples of backreaction of small scale inhomogeneities in cosmology
}

\author{Stephen R. Green}
\email{sgreen04@uoguelph.ca}
\thanks{CITA National Fellow}
\affiliation{Department of Physics \\
  University of Guelph \\
   Guelph, Ontario N1G 2W1, Canada}
\author{Robert M. Wald}
\email{rmwa@uchicago.edu}
\affiliation{Enrico Fermi Institute and Department of Physics \\
  University of Chicago \\
  5620 South Ellis Avenue, Chicago, Illinois 60637, U.S.A.}

\date{\today}

\begin{abstract}

  In previous work, we introduced a new framework to treat large scale
  backreaction effects due to small scale inhomogeneities in general
  relativity.  We considered one-parameter families of spacetimes for
  which such backreaction effects can occur, and we proved that,
  provided the weak energy condition on matter is satisfied, the
  leading effect of small scale inhomogeneities on large scale
  dynamics is to produce a traceless effective stress-energy tensor
  that itself satisfies the weak energy condition. In this work, we
  illustrate the nature of our framework by providing two explicit
  examples of one-parameter families with backreaction.  The first,
  based on previous work of Berger, is a family of polarized vacuum
  Gowdy spacetimes on a torus, which satisfies all of the assumptions
  of our framework.  As the parameter approaches its limiting value,
  the metric uniformly approaches a smooth background metric, but
  spacetime derivatives of the deviation of the metric from the
  background metric do not converge uniformly to zero. The limiting
  metric has nontrivial backreaction from the small scale
  inhomogeneities, with an effective stress-energy that is traceless
  and satisfies the weak energy condition, in accord with our
  theorems.  Our second one-parameter family consists of metrics which
  have a uniform Friedmann-Lema\^itre-Robertson-Walker limit. This
  family satisfies all of our assumptions with the exception of the
  weak energy condition for matter.  In this case, the limiting metric
  has an effective stress-energy tensor which is not traceless.  We
  emphasize the importance of imposing energy conditions on matter in
  studies of backreaction.

\end{abstract}

\maketitle

\section{Introduction}

The standard model of cosmology is based on a hierarchy of length
scales.  At the largest scales in cosmology, it is generally believed
that the universe is well-described by a metric with
Friedmann-Lema\^itre-Robertson-Walker (FLRW) symmetry that satisfies
Einstein's equation with source given by the averaged stress-energy of
matter.  At scales somewhat smaller than this, the deviations from the
FLRW solution are believed to be well described by linear perturbation
theory.  Finally, below a certain scale, nonlinear effects become
important, and numerical $N$-body simulations are normally used to make
predictions. The above beliefs underlie the standard cosmological
model, which has proven to be extremely successful to date.

Despite the success of the standard cosmological model, objections
have been raised that nonlinear effects due to the small scale structure
are not being fully taken into account (see, e.g., \cite{Ellis:1984}).
The Friedmann equations neglect averages of terms
nonlinear in small scale perturbations, which are present in the
Einstein equation.  Since nonlinearities are important for small scale
dynamics, how do we know that their averaged effects cannot contribute
at large scales? Might it even be possible that such
nonlinearities could produce effects on large scale dynamics
that mimic the effect of a cosmological constant?

At small scales, where nonlinear dynamics are important, fractional
density perturbations can be much larger than 1. This formed the basis
of Ellis's original argument \cite{Ellis:1984} that neglecting
nonlinear terms in the Einstein equation could be problematic.
However, Ishibashi and Wald \cite{Ishibashi:2005sj} argued that, even
though perturbations in the matter stress-energy can be large,
perturbations in the metric (as opposed to derivatives of the metric)
should nevertheless be small, provided that appropriate gauge choices
are made\footnote{Other authors continue to claim that the metric
  perturbation cannot be made small globally
  \cite{Ellis:2011hk}.}. They argued that, except in the immediate
vicinity of strong field objects such as black holes and neutron
stars, our universe should be well described on all scales by a
Newtonianly perturbed FLRW metric, and that the terms in Einstein's
equation that are nonlinear in the deviation of the metric from an
FLRW metric should be negligible.

However, it is clear that, in general, non-Newtonian circumstances,
even when the deviation of the metric from a metric with FLRW symmetry
is small, the nonlinear terms in the deviation of the metric from FLRW
symmetry {\em can} affect the large scale dynamics. Indeed, if the
universe were filled with low amplitude, high frequency gravitational
radiation, then the metric could be uniformly close to a metric with
FLRW symmetry, but this FLRW metric would behave dynamically as though
the universe were filled with a $P=\frac{1}{3}\rho$ fluid.  This
dynamical behavior occurs as a consequence of the nonlinear terms in
Einstein's equation.  Thus, an interesting issue in general
relativity is to quantify the degree to which small scale spacetime
inhomogeneities can, through nonlinear interactions, affect large
scale dynamics.

Prior to our work \cite{Green:2010qy}, a number of
averaging formalisms had been developed to try to address this issue.
The most widely studied approach was initiated by Buchert
\cite{Buchert:1999er,Buchert:2001sa}.  Here, one works in comoving,
synchronous coordinates, so that the metric takes the form
\begin{equation}
  ds^2 = - dt^2 + q_{ij}(t,x) dx^i dx^j.
\label{syn}
\end{equation}
One can then define averages of scalars over (compact, comoving regions of)
constant-$t$ hypersurfaces.  In particular, Buchert defines an
averaged ``scale factor'' as the cube root of the volume of one of
these regions. Equations of motion for the averaged
quantities are obtained by averaging (components of) Einstein's
equation. In particular, by averaging the Hamiltonian constraint and
the Raychaudhuri equation, analogs of the Friedmann equations can be
derived in this framework.

There are two independent serious difficulties with this
framework. First, it is far from clear how to physically interpret the
averaged quantities defined in this framework. For example, if the
``scale factor'' increases with time, this does not mean that
observers would see anything like Hubble's law holding. Furthermore,
small changes to the spacetime metric can produce significant changes
to the behavior of geodesics and, consequently, significant changes to
quantities such as the time dependence of volumes of comoving regions,
without there being any correspondingly large physical effects.  To
see this more concretely, suppose that one were interested in
estimating the nonlinear backreaction effects produced by the planets
on the metric of the solar system. One way of doing this would be to
use ordinary perturbation theory about the stationary metric of the
sun. If one calculates the linearized perturbations produced by the
planets and estimates their second and higher order corrections, one
would correctly conclude that the nonlinear backreaction effects of
the planets on the metric of the solar system are extremely small. But
another way of estimating the nonlinear backreaction effects of the
planets---which is closely analogous to the Buchert approach in
cosmology---would be to sprinkle some irrotational dust throughout the
solar system and investigate the effects of the planets on the motion
of this dust. The comoving dynamics of the dust would be described by
a synchronous metric of the form \eqref{syn}. One would find that the
dust geodesics near the planets would immediately begin shearing; the
nonlinear effects of shear would soon result in significant
convergence and eventually there would be caustics. The dynamics of
the dust would be quite complicated and would give rise to large
nonlinear effects on the metric \eqref{syn} attributable to the
planets.  One might then be tempted to conclude that nonlinear effects
produced by the planets result in large backreaction effects on the
spacetime metric of the solar system. Obviously, this is not the case:
These ``large effects'' are gauge artifacts resulting from the use of
synchronous coordinates rather than coordinates in which the metric is
(nearly) stationary.

A second serious difficulty of this approach is that the ``Friedmann
equations'' derived in this framework are not closed. They contain a
scalar quantity $\mathcal{Q}_\mathcal{D}$, arising from averaging
nonlinear terms in the Einstein equation, which is called the {\em
  kinematical backreaction} scalar. The evolution of
$\mathcal{Q}_\mathcal{D}$ is undetermined in the framework.  This is
because the framework considers only the spatial average of two
components of Einstein's equation, and thus is discarding much of the
content of Einstein's equation.

Another approach to backreaction, known as {\em macroscopic gravity}, 
was initiated by Zalaletdinov \cite{Zalaletdinov:1996aj}.
Rather than restricting to averaging scalar quantities on spatial
slices, in macroscopic gravity, an averaging operator $\langle\cdot\rangle$ is
introduced which allows one to define local spacetime averages of
tensor quantities.  The idea is to derive an equation for the average
$\langle g_{ab}\rangle$ of the actual spacetime metric $g_{ab}$.
Backreaction in this framework occurs through the ``connection
correlation tensor'',
\begin{equation}\label{eq:Z}
Z_{\phantom{a}bc\phantom{d}ef}^{a\phantom{bc}d} = \langle\Gamma^a_{\phantom{a}b[c}\Gamma^d_{\phantom{d}|e|f]}\rangle - \langle\Gamma^a_{\phantom{a}b[c}\rangle\langle\Gamma^d_{\phantom{d}|e|f]}\rangle,
\end{equation}
where $\Gamma^a_{\phantom{a}bc}$ is the Christoffel connection.
Various properties are derived for $Z_{abcdef}$, but, as with
$\mathcal{Q}_{\mathcal{D}}$, the full content of Einstein's equation is not used.

Because of the presence of undetermined quantities in the averaged
frameworks of Buchert and Zalaletdinov, a common approach to ``solve''
the equations has been to {\em assume} a particular form for the
backreaction tensors, and from that starting point, derive results for
the large scale behavior \cite{Coley:2005ei,Buchert:2010ug}.  However,
no argument is given that the assumed form of the backreaction tensors
can actually arise from averaging a physically reasonable small scale
matter distribution.  For example, in \cite{Buchert:2010ug}, Buchert
and Obadia assume that $\mathcal{Q}_{\mathcal{D}}$ is equivalent to a
scalar field (called the {\em morphon field} \cite{Buchert:2006ya}),
with a particular potential.  With a suitable choice of potential,
they find that inflation can occur in vacuum spacetimes due to the
presence of inhomogeneities.  However, one is not free to
arbitrarily specify the large scale effects of backreaction: They must
be shown to arise from averaging an actual inhomogeneous spacetime.

Our approach---which we summarize in detail in section
Sec.~\ref{sec:framework}---bears considerable resemblance to that of
Zalaletdinov \cite{Zalaletdinov:1996aj} in that we define a
backreaction tensor $\mu_{abcdef}$, which is closely related to
Zalaletdinov's $Z_{abcdef}$.  However, we make full use of Einstein's
equation, and we have better mathematical control over the
perturbations through our introduction of one-parameter families,
which allow us to rigorously take a short wavelength
limit. ``Averaging'' of nonlinear terms in Einstein's equation may
then be rigorously defined via the use of weak limits.  We were thus
able to derive strong constraints on $\mu_{abcdef}$, and we proved the
following theorem about dynamical behavior in the limit of short
wavelength perturbations: If the matter stress-energy tensor satisfies
the weak energy condition (i.e., has positive energy density, as
measured by any timelike observer), then the leading order effect of
small scale inhomogeneities on the background metric is to produce an
effective stress-energy tensor that is traceless and itself satisfies
the weak energy condition.  In particular, backreaction produced by
small scale inhomogeneities cannot mimic a cosmological constant.

However, our work merely assumed one-parameter families with the
properties stated in Sec.~\ref{sec:framework} below; we did not prove
existence of a family with nontrivial backreaction, thus leaving open
the possibility that our assumptions are self-consistent only in the
case of no backreaction.  Objections also have been raised that the
interpretation of the parameter appearing in our one-parameter
families is unclear \cite{Rasanen:2011bm}, and that our averaging
scheme is ``ultralocal'', and thus cannot capture effects due to
matter inhomogeneities over finite regions \cite{Clarkson:2011zq}.
Therefore, it seems clear that it would be useful for us to provide a
concrete example of a one-parameter family of spacetimes with
nontrivial backreaction that satisfies our assumptions. Such an
example would serve the multiple purposes of proving the
self-consistency of our assumptions, illustrating the meaning of the
parameter appearing in our one-parameter families, and showing
explicitly how our theorems have direct implications for perturbations
of finite amplitude and wavelength.

The purpose of this paper is to provide two examples of one-parameter
families of spacetimes with nontrivial backreaction. The first,
analyzed in Sec.~\ref{sec:vacuum}, is adapted from \cite{Berger:1974}
and is a family of vacuum Gowdy spacetimes. This family satisfies all
of the assumptions of our work and has nontrivial backreaction in that
the limiting metric does not satisfy the vacuum Einstein equation.
Its effective stress-energy tensor can be explicitly seen to be
traceless and satisfy the weak energy condition, as guaranteed by the
theorems in \cite{Green:2010qy}.

The second family, which we produce in Sec.~\ref{sec:synge}, is
obtained by ``Synge's method,'' i.e., by simply writing down a family
of metrics and declaring the matter stress-energy of each spacetime to
be equal to the Einstein tensor of the metric, thereby trivially
``solving'' Einstein's equation \cite{SyngesMethod}.  This family is
non-vacuum and approaches a FLRW spacetime as the parameter approaches
its limiting value. It satisfies all of our assumptions except that
the matter stress-energy does not satisfy the weak energy condition.
We obtain a nontrivial effective stress-energy tensor, which does not
satisfy the properties of our theorems.  This example shows that if
matter violates the weak energy condition (or if Einstein's equation
is not sufficiently utilized), then it is easy to produce spacetimes
with significant backreaction that is unconstrained by our theorems.
This illustrates the necessity of applying the full content of
Einstein's equation with physically reasonable energy conditions
imposed upon matter in any analysis of backreaction.

In this work, we follow all notation and sign conventions of \cite{Wald:1984}.

\section{Framework}\label{sec:framework}

In this section we review our framework and we summarize the main
results of \cite{Green:2010qy}. Our framework is a generalization to the
non-vacuum case of a framework proposed by Burnett \cite{Burnett:1989gp}, 
which itself is a rigorous version of Isaacson's approach
\cite{Isaacson:1967zz,*Isaacson:1968zza}.

As indicated in the introduction, to analyze backreaction, we consider
a one-paramater family of spacetimes $g_{ab}(\lambda)$ for
$0\le\lambda\le\lambda_0$.  We require $g_{ab}(\lambda)$ to be a
solution to Einstein's equation for all $\lambda>0$, and, as
$\lambda\to0$, we require uniform convergence of $g_{ab}(\lambda)$ to
a ``background metric'' $g_{ab}^{(0)}\equiv g_{ab}(0)$.  One-parameter
families of metrics are also employed in this manner to properly treat
ordinary perturbation theory (see, e.g., Sec.~7.5 of
\cite{Wald:1984}).  However, whereas in ordinary perturbation theory
the metric is required to be jointly smooth in $\lambda$ and the
coordinates $x$, here we require only that spacetime derivatives of
$(g_{ab}(\lambda)-g_{ab}^{(0)})$ are {\em bounded} as $\lambda\to0$.
This elevates the importance of derivatives of perturbations of the
background metric so that, {\em a priori}, small scale inhomogeneities
can play a dynamical role in the evolution of the background metric.
Indeed, in contrast to ordinary perturbation theory, it is not true,
in general, that the Einstein tensor has a uniform limit as
$\lambda\to0$, and it does not follow that $g_{ab}^{(0)}$ is a
solution to the Einstein equation.  However, if we add a suitable
assumption that ``spacetime averages exist'', then we can derive an
equation for $g_{ab}^{(0)}$, which contains the contributions arising
from the backreaction produced by small scale inhomogeneities.

Our precise assumptions are as follows.  Fix a spacetime manifold $M$
with derivative operator $\nabla_a$, and let $g_{ab}(\lambda)$ be a
one-parameter family of metrics on $M$, defined for $\lambda\ge 0$.
Let $e_{ab}$ be an arbitrary Riemannian metric on $M$, and define
$|t_{a_1\dots a_n}|^2=e^{a_1b_1}\dots e^{a_nb_n}t_{a_1\dots
  a_n}t_{b_1\dots b_n}$.  Suppose now that the following assumptions
hold:
\renewcommand{\theenumi}{\roman{enumi}}
\renewcommand{\labelenumi}{(\theenumi)}
\begin{enumerate}
\item{\label{assumption1} For all $\lambda>0$, we have 
\begin{equation}
G_{ab}(g(\lambda)) + \Lambda g_{ab}(\lambda) = 8 \pi T_{ab}(\lambda), 
\label{Ee}
\end{equation}
where $T_{ab}(\lambda)$ satisfies the weak energy condition,
i.e., for all $\lambda >0$ we have \mbox{$T_{ab}(\lambda) t^a(\lambda)
  t^b (\lambda) \geq 0$} for all vectors $t^a(\lambda)$ that are
timelike with respect to $g_{ab}(\lambda)$.}
\item{\label{assumption2}There exists a smooth positive function $C_1(x)$ on $M$ such that 
\begin{equation}
|h_{ab}(\lambda,x)| \leq \lambda C_1(x),
\end{equation}}
where $h_{ab} (\lambda,x) \equiv g_{ab}(\lambda,x) - g_{ab}(0,x)$.
\item{\label{assumption3}There exists a smooth positive function $C_2(x)$ on $M$ such that 
\begin{equation}
|\nabla_c h_{ab}(\lambda,x)| \leq C_2(x).
\end{equation}}
\item{\label{assumption4}There exists a smooth tensor field $\mu_{abcdef}$ on $M$ such that 
\begin{equation}
\wlim_{\lambda \rightarrow 0}
\left[\nabla_ah_{cd}(\lambda)\nabla_bh_{ef}(\lambda) \right] =
\mu_{abcdef},
\label{mu}
\end{equation}
where ``$\wlim$'' denotes the weak limit.}
\end{enumerate}
\renewcommand{\labelenumi}{\theenumi}

The notion of ``weak limit'', which appears in assumption (\ref{assumption4}),
corresponds roughly to taking a local spacetime average, and then
taking the limit as $\lambda\to0$.  More precisely, $A_{a_1\dots
  a_n}(\lambda)$ converges {\em weakly} to $A_{a_1\dots a_n}^{(0)}$ as
$\lambda\to0$ if and only if, for all smooth tensor fields
$f^{a_1\dots a_n}$ of compact support,
\begin{equation}
  \lim_{\lambda\to0}\int f^{a_1\dots a_n}A_{a_1\dots a_n}(\lambda)=\int f^{a_1\dots a_n}A_{a_1\dots a_n}^{(0)}.
\end{equation}

Assumptions (\ref{assumption1})--(\ref{assumption4}) allow us to
derive an equation satisfied\footnote{Note that assumption
(\ref{assumption1}) only requires that the Einstein equation hold for
$\lambda>0$, but not for $\lambda=0$.} by $g_{ab}^{(0)}$. Indeed, in \cite{Green:2010qy}
we showed that $g_{ab}(0)$ satisfies the equation
\begin{equation}\label{Eeg0}
  G_{ab}(g^{(0)})+\Lambda g_{ab}^{(0)}=8\pi T^{(0)}_{ab} + 8\pi t_{ab}^{(0)}.
\end{equation}
Here, $T_{ab}^{(0)}\equiv\wlim_{\lambda\to0}T_{ab}(\lambda)$---which
must exist as a result of the assumptions---and $t_{ab}^{(0)}$ is a
particular linear combination of contractions of the tensor
$\mu_{abcdef}$.  The tensor $t_{ab}^{(0)}$ is called the ``effective
gravitational stress-energy tensor,'' and it describes the
dominant\footnote{\label{fn:higherorder}Higher order contributions to
  back reaction were also derived in \cite{Green:2010qy}. In
  \cite{Green:2011wc}, we showed that the dominant effect of the
  higher order contributions on the dynamics of our universe is to
  modify the Friedmann equations by the inclusion of effective
  stress-energy contributions associated with Newtonian gravitational
  potential energy and stresses and with the kinetic motion of matter
  (see also \cite{Futamase:1996fk,Baumann:2010tm}).} contribution to
backreaction due to small scale inhomogeneities.

In \cite{Green:2010qy}, we proved two theorems constraining $t_{ab}^{(0)}$:
\begin{theorem}\label{thm1}Given a one-parameter family $g_{ab}(\lambda)$
  satisfying assumptions (\ref{assumption1})--(\ref{assumption4})
  above, the effective stress-energy tensor $t^{(0)}_{ab}$ appearing
  in equation \eqref{Eeg0} for the background metric $g^{(0)}_{ab}$ is
  traceless,
\begin{equation}
{t^{(0)a}}_a = 0.
\end{equation}
\end{theorem}
\begin{theorem}\label{thm2}Given a one-parameter family $g_{ab}(\lambda)$
  satisfying assumptions (\ref{assumption1})--(\ref{assumption4})
  above, the effective stress-energy tensor $t^{(0)}_{ab}$ appearing
  in equation \eqref{Eeg0} for the background metric $g^{(0)}_{ab}$
  satisfies the weak energy condition, i.e.,
\begin{equation}
t^{(0)}_{ab} t^a t^b \geq 0 
\end{equation}
for all $t^a$ that are timelike with respect to $g^{(0)}_{ab}$.
\end{theorem}

In essence, these theorems show that only those small scale metric
inhomogeneities corresponding to gravitational radiation can have a
significant backreaction effect\footnote{Although matter
  inhomogeneities can produce only a negligibly small effect on the
  {\it dynamical evolution} of the background FLRW metric, they can
  produce large observable effects on the {\it apparent luminosity} of
  objects in the universe via gravitational lensing; see, e.g.,
  \cite{Holz:1997ic}.}. In particular, in the case where we have FLRW
symmetry, the effective stress-energy tensor must be of the form of a
$P=\frac{1}{3}\rho$ fluid, and therefore cannot mimic dark energy.

Finally, we note that in the statement of assumptions
(\ref{assumption1})--(\ref{assumption4}) above, $\lambda$ is a
``continuous parameter,'' i.e., it takes all real values in an
interval $[0, \lambda_0]$ for some $\lambda_0 > 0$.  However, there
would be no essential change if we took $\lambda$ to be a ``discrete
parameter'' in our assumptions---such as $\lambda = \{1/N\}$ for all
positive integers $N$---provided only that $\lambda$ takes positive
values that limit to $0$. It will be convenient to use such a discrete
parameter in the example of the next section.

\section{Backreaction in vacuum Gowdy spacetimes}\label{sec:vacuum}

Gowdy spacetimes are exact plane-symmetric vacuum solutions which
describe closed cosmologies containing gravitational waves
\cite{Gowdy:1971jh,*Gowdy:1973mu}.  We restrict to the case where
spatial slices have topology $T^3$.
The general $T^3$ Gowdy metric may be written in the form \cite{Ringstrom:2002km}
\begin{equation}
  ds^2_{\text{Gowdy}} = e^{(\tau-\alpha)/2}\left(-e^{-2\tau}d\tau^2 + d\theta^2\right) + e^{-\tau}\left[ e^P d\sigma^2 + 2 e^P Q d\sigma d\delta + \left( e^P Q^2 + e^{-P}\right) d\delta^2\right],
\end{equation}
where the functions $\alpha$, $P$ and $Q$ depend only on $\tau$ and
$\theta$ and the spatial coordinates have the range
$0\le\theta,\sigma,\delta<2\pi$, with periodic boundary conditions.
We will restrict consideration to the case $Q=0$,
known as the {\em polarized} Gowdy spacetimes.

When $Q=0$, the vacuum Einstein equations reduce to the evolution equation,
\begin{equation}
  \label{eq:Ppol}0=\ddot{P} - e^{-2\tau}P'',
\end{equation}
and the constraints,
\begin{align}
 \label{eq:const1} \dot{\alpha} &= \dot{P}^2 + e^{-2\tau}(P')^2 \, ,\\
 \label{eq:const2}\alpha'&=2 P'\dot{P} \, .
\end{align}
Here, the prime and dot denote differentiation with respect to
$\theta$ and $\tau$, respectively.
The general
solution to \eqref{eq:Ppol} is
\begin{equation}
  P=A_0+B_0\tau+\sum_{n=1}^\infty\left[ A_n J_0\left(n e^{-\tau}\right) + B_n Y_0\left(ne^{-\tau}\right)\right]\sin\left(n\theta+\phi_n\right),
\end{equation}
where $A_n$, $B_n$, and $\phi_n$ are free parameters.  The discrete
index $n$ arises from the compactness of the $\theta$ direction.

Following \cite{Berger:1974} (see also
\cite{Charach:1978vf}), we now construct a one-parameter family satisfying
assumptions (\ref{assumption1})--(\ref{assumption4}). 
As discussed in the last paragraph of the previous section,
our family will be parametrized by a discrete
parameter $N$, and the limit $N\to\infty$ will correspond to $\lambda\to0$.
We choose 
\begin{equation}
  P_N=\frac{A}{\sqrt{N}}J_0\left(Ne^{-\tau}\right)\sin(N\theta),
\end{equation}
where $A$ is an arbitrary constant.
The constraint equations may be solved by direct integration
to yield
\begin{align}
  \alpha_N ={}& -\frac{A^2e^{-\tau}}{2}J_1\left(Ne^{-\tau}\right)J_0\left(Ne^{-\tau}\right)\cos\left(2N\theta\right)\nonumber\\
  & -\frac{A^2Ne^{-2\tau}}{4}\left\{\left[J_0\left(Ne^{-\tau}\right)\right]^2+2\left[J_1\left(Ne^{-\tau}\right)\right]^2-J_0\left(Ne^{-\tau}\right)J_2\left(Ne^{-\tau}\right)\right\}.
\end{align}

From the asymptotic form of 
the Bessel function, we find that for large values of $N$,
\begin{equation}
  P_N\to \frac{A}{N} \sqrt{\frac{2e^\tau}{\pi}}\cos\left(Ne^{-\tau}-\frac{\pi}{4}\right)\sin\left(N\theta\right).
\end{equation}
From this and the similar asymptotic form of $\alpha_N$, it may be
verified that assumptions
\mbox{(\ref{assumption1})--(\ref{assumption4})} of the previous
section hold for the family $\{P_N,\alpha_N\}$.  In particular, as
$N\to\infty$, we have
\begin{align}
 \lim_{N\to\infty} P_N &= 0,\\
  \lim_{N\to\infty} \alpha_N &= -\frac{A^2e^{-\tau}}{\pi}
\end{align}
where the convergence is uniform on compact sets.
Thus, the ``background metric'' is
\begin{equation}\label{eq:gowdybg}
  ds^2_{(0)}=e^{\left(\tau+A^2e^{-\tau}/\pi\right)/2}\left(-e^{-2\tau}d\tau^2+d\theta^2\right)+e^{-\tau}\left(d\sigma^2+d\delta^2\right).
\end{equation}

Despite the fact that for each $N<\infty$, the metric is a vacuum
solution, the limiting metric \eqref{eq:gowdybg} is {\em not} a vacuum solution.
The effective stress-energy, $t_{ab}^{(0)}$, could be computed by computing 
$\mu_{abcdef}$ via \eqref{mu} and using the formula for 
$t_{ab}^{(0)}$ in terms of $\mu_{abcdef}$ given in \cite{Green:2010qy}.
However, it is far easier to obtain $t_{ab}^{(0)}$ by simply computing
the left side of \eqref{Eeg0}. It is easily seen that the 
nonvanishing components of the Einstein tensor
of the metric \eqref{eq:gowdybg} are given by
\begin{align}
  G_{\tau\tau}\left(g^{(0)}\right)&=\frac{A^2e^{-\tau}}{4\pi} \nonumber \\
  G_{\theta\theta}\left(g^{(0)}\right)&=\frac{A^2e^{\tau}}{4\pi}.
\label{Gab}
\end{align}
Note that the effective stress-energy 
$t_{ab}^{(0)} = G_{ab}/8 \pi$
associated with \eqref{Gab} is
traceless and satisfies the weak energy condition, in accord with
our general theorems.

Thus, we have provided an explicit example of a one-parameter 
family wherein the small scale inhomogeneities 
produce a nontrivial backreaction. One can see explicitly in this example
how, in our framework, the vacuum, inhomogeneous spacetime 
metric $g_{ab}(N)$ can be well approximated for large but finite $N$ 
by the non-vacuum, homogeneous spacetime metric
\eqref{eq:gowdybg}.

\section{Backreaction without energy conditions}\label{sec:synge}

In this section, we provide an example of a one-parameter family that
satisfies assumptions (\ref{assumption2})--(\ref{assumption4}) of
Sec.~\ref{sec:framework} but does not satisfy the requirement of
assumption (\ref{assumption1}) that the matter stress-energy tensor
satisfy the weak energy condition. This illustrates the importance of
the weak energy condition on matter for the validity of Theorems
\ref{thm1} and \ref{thm2} of Sec.~\ref{sec:framework}.

For our one-parameter family, $g_{ab}(\lambda)$,
we choose metrics conformally related to a spatially flat FLRW
metric $g_{ab}^{(0)}$, i.e.,
\begin{equation}
  g_{ab}(\lambda) = \Omega^2(\lambda)g_{ab}^{(0)}=\Omega^2(\lambda)a^2(\tau)\eta_{ab},
\label{cflrw}
\end{equation}
where $a(\tau)$ is chosen arbitrarily.
We choose the conformal factor to be
\begin{equation}
  \log\Omega(\lambda) = \lambda A\left[\sin\left(\frac{x}{\lambda}\right)+\sin\left(\frac{y}{\lambda}\right)+\sin\left(\frac{z}{\lambda}\right)\right],
\label{Omega}
\end{equation}
with $A$ constant.  Of course, $g_{ab}(\lambda)$ does not satisfy
Einstein's equation with any known form of matter.
Nevertheless, we may simply declare the existence of a new form
of matter with stress-energy tensor given by 
\begin{equation}
  T_{ab}(\lambda) = \frac{1}{8\pi}G_{ab}(g(\lambda)) \, 
\end{equation}
for all $\lambda > 0$.  Then $g_{ab}(\lambda)$ is a solution of
Einstein's equation (with vanishing cosmological constant, $\Lambda =
0$) for all $\lambda > 0$. This procedure for ``solving'' Einstein's
equation is usually referred to as ``Synge's method''
\cite{SyngesMethod}.  It can then be easily verified that our
one-parameter family \eqref{cflrw} and \eqref{Omega} satisfies all of
the assumptions of Sec.~\ref{sec:framework} except that the
stress-energy tensor of our new form of matter does not satisfy the
weak energy condition for $\lambda > 0$.

By \eqref{Eeg0},
the effective stress-energy tensor produced by the small scale inhomogeneities is
\begin{align}\label{eq:syngelimit}
  t_{ab}^{(0)} ={}& \frac{1}{8\pi}G_{ab}(g^{(0)})-T_{ab}^{(0)}\nonumber\\
  ={}& \frac{1}{8\pi}\wlim_{\lambda\to0}\left[G_{ab}(g^{(0)})-G_{ab}(g(\lambda))\right]\nonumber\\
  ={}& \frac{1}{8\pi} \wlim_{\lambda\to0}\left[ 2\nabla_a\nabla_b\log\Omega - 2 g_{ab}^{(0)}g^{(0)cd}\nabla_c\nabla_d\log\Omega \right.\nonumber\\
  & \left. -2\left(\nabla_a\log\Omega\right)\left(\nabla_b\log\Omega\right) - g^{(0)}_{ab} g^{(0)cd}\left(\nabla_c\log\Omega\right)\left(\nabla_d\log\Omega\right)\right].
\end{align}
Here, on the last line, we used the expression relating the two Ricci tensors (see Appendix D of \cite{Wald:1984}),
\begin{align}\label{eq:riccirelation}
  R_{ab}(g(\lambda)) ={}& R_{ab}(g^{(0)}) - 2 \nabla_a\nabla_b\log\Omega - g_{ab}^{(0)}g^{(0)cd}\nabla_c\nabla_d\log\Omega\nonumber\\
  &+2\left(\nabla_a\log\Omega\right)\left(\nabla_b\log\Omega\right) - 2 g^{(0)}_{ab} g^{(0)cd}\left(\nabla_c\log\Omega\right)\left(\nabla_d\log\Omega\right),
\end{align}
where $\nabla_a$ is the derivative operator associated with
$g_{ab}^{(0)}$.  A straightforward calculation\footnote{The only
  nontrivial weak limits involved are of the form
  $\wlim_{\lambda\to0}\cos^2(x/\lambda)=1/2$.} yields
\begin{align}
  t^{(0)}_{00}&=\frac{3}{16\pi}A^2 \\
  t^{(0)}_{0i}&= 0 \, \\
  t^{(0)}_{ij}&=-\frac{5}{16\pi}A^2\delta_{ij} \, .
\end{align}

Thus, our effective stress-energy tensor produced by small-scale
inhomogeneities corresponds to a $P = -\frac{5}{3} \rho$ fluid. This
fails\footnote{Note also that $t_{ab}^{(0)}$ and $T_{ab}^{(0)}$ are
  not separately conserved with respect to the background derivative
  operator, although, of course, by the Bianchi identity for
  $G_{ab}(g^{(0)})$, we have $\nabla^b(T_{ab}^{(0)}+t_{ab}^{(0)}) =
  0$.} to be traceless and fails to satisfy the weak energy condition,
in violation of both Theorems \ref{thm1} and \ref{thm2} of
Sec.~\ref{sec:framework}.  This should come as no surprise, since we
have failed to satisfy assumption (\ref{assumption1}).  This example
thus shows that assumption (\ref{assumption1}) is essentially needed
for the validity of our main results.

We note that several analyses \cite{Coley:2005ei,Coley:2006kp} within
the general framework of \cite{Zalaletdinov:1996aj} have found an
effective $P=-\frac{1}{3}\rho$ fluid\footnote{Due to the nature of the
  constructions, these analyses are guaranteed to find an effective
  stress-energy tensor of FLRW form, with components
  $t_{\mu\nu}^{(0)}$ which are constant.  When one further imposes
  that this effective stress-energy tensor be conserved, a
  $P=-\frac{1}{3}\rho$ fluid is automatic.}. As in our example above,
this result is inconsistent\footnote{We remind the reader that
  additional effects are possible at higher order in perturbation
  theory (see footnote \ref{fn:higherorder}).} with the tracelessness
of $t_{ab}^{(0)}$ (Theorem \ref{thm1} of Sec.~\ref{sec:framework}). We
believe that the violation of the tracelessness of $t_{ab}^{(0)}$ in
\cite{Coley:2005ei,Coley:2006kp} is of a similar origin: As discussed
in the introduction, the full content of Einstein's equation has not
been utilized, and the requirement that matter satisfy the weak energy
condition has not been imposed.

\section{Conclusions}

We have provided two examples of one-parameter families of
cosmological spacetimes with nontrivial backreaction.  The first
describes a universe filled with gravitational radiation, and it
exhibits a traceless effective stress-energy tensor satisfying the
weak energy condition, consistent with our theorems.  The second
example contains matter which violates the weak energy condition (a
key assumption necessary to prove our theorems) and, as a consequence,
contains a more exotic $P=-\frac{5}{3}\rho$ effective fluid.

These examples provide insight into our framework for treating
backreaction \cite{Green:2010qy}, as well as the averaging problem in
general.  In particular, we conclude that the appearance of effective
stress-energies violating our theorems in other approaches to
backreaction is closely related to the underdetermination of the
equations of motion, which in turn is closely related to a failure to
properly impose local energy conditions on matter.

\begin{acknowledgments}

  We wish to thank Hans Ringstr\"om and Alan Rendall for helpful
  discussions. This research was supported in part by NSF grants
  PHY~08-54807 and PHY~12-02718 to the University of Chicago, and by
  NSERC.  SRG is supported by a CITA National Fellowship at the
  University of Guelph, and he thanks the Perimeter Institute for
  hospitality.

\end{acknowledgments}

\bibliography{mybib.bib}

\end{document}